\documentclass[aps,prd,11pt,showpacs,showkeys,preprintnumbers,superscriptaddress,nobibnotes,floatfix,longbibliography,nofootinbib]{revtex4-1}

\pdfoutput=1
\usepackage{mathrsfs}
\usepackage{hyperref}
\usepackage{graphicx}
\usepackage{amsfonts,amsmath,amssymb,bm}
\usepackage{array}
\usepackage{color}
\usepackage{enumitem}
\usepackage[explicit]{titlesec}
\usepackage[utf8]{inputenc}
\usepackage{graphicx}
\usepackage{multirow}
\usepackage{lineno}
\usepackage{ulem}
\usepackage[caption=false]{subfig}
\usepackage{upgreek}
\usepackage{threeparttable}

\def\tabnotefont{\fontsize{9}{10}\selectfont}%

\begin{document}

\title{A practical approach of measuring $^{238}$U and $^{232}$Th in liquid scintillator to sub-ppq level using ICP-MS}

\author{Yuanxia Li}
\affiliation{Institute of High Energy Physics, Chinese Academy of Sciences, Beijing, China}
\author{Jie Zhao}
\email{zhaojie@ihep.ac.cn}
\affiliation{Institute of High Energy Physics, Chinese Academy of Sciences, Beijing, China}
\author{Yayun Ding}
\affiliation{Institute of High Energy Physics, Chinese Academy of Sciences, Beijing, China}
\author{Tao Hu}
\affiliation{Institute of High Energy Physics, Chinese Academy of Sciences, Beijing, China}
\author{Jiaxuan Ye}
\affiliation{Institute of High Energy Physics, Chinese Academy of Sciences, Beijing, China}
\author{Jian Fang}
\affiliation{Institute of High Energy Physics, Chinese Academy of Sciences, Beijing, China}
\author{Liangjian Wen}
\affiliation{Institute of High Energy Physics, Chinese Academy of Sciences, Beijing, China}

\begin{abstract}
Liquid scintillator (LS) is commonly utilized in experiments seeking rare events due to its high light yield, transparency, and radiopurity. The concentration of $^{238}$U and $^{232}$Th in LS consistently remains below 1 ppq (10$^{-15}$~g/g), and the current screening result is based on a minimum 20-ton detector. Inductively coupled plasma mass (ICP-MS) spectroscopy is well-regarded for its high sensitivity to trace $^{238}$U and $^{232}$Th. This study outlines a method for detecting $^{238}$U and $^{232}$Th in LS at the sub-ppq level using ICP-MS, involving the enrichment of $^{238}$U/$^{232}$Th from the LS through acid extraction. With meticulous cleanliness control, $^{238}$U/$^{232}$Th in approximately 2~kg of LS is concentrated by acid extraction with 0.4 (0.3)~pg $^{238}$U ($^{232}$Th) contamination. Three standard adding methods are employed to assess recovery efficiency, including radon daughter, 2,5-diphenyloxazole (PPO), and natural non-existent $^{233}$U/$^{229}$Th. The method detection limit at a 99\% confidence level of this approach can reach approximately 0.2-0.3~ppq for $^{238}$U/$^{232}$Th with nearly 100\% recovery efficiency.
\end{abstract}

\maketitle
\date{\today}

\section{Introduction}
Liquid scintillator (LS) plays a crucial role in the detection and measurement of radiation, making it an essential tool in various scientific and industrial applications. In high-energy physics experiments, LS is used in particle detectors to measure the energy and trajectory of subatomic particles produced in collisions, helping scientists understand the fundamental properties of matter and the universe. The pursuit of rare events demands that the radiopurity of $^{238}$U/$^{232}$Th in LS for neutrino experiments be kept below sub-ppq (10$^{-15}$ g/g) levels, as shown in Table~\ref{tab:LSexperiment}. Achieving such stringent radiopurity standards renders the screening process exceptionally challenging. The measurement results in Table~\ref{tab:LSexperiment} are all based on measurements of the $^{238}$U/$^{232}$Th content in the LS after the detector has been filled, with detector sizes ranging from tens to hundreds of tons. Currently, there are no experiments that can measure the $^{238}$U/$^{232}$Th content in the LS at sub-ppq levels based on laboratory samples in the kilogram range.

\begin{table}[!h]
\begin{center}
\caption{The table lists the $^{238}$U/$^{232}$Th content in the target media of large-scale LS experiments worldwide. Only the Daya Bay experiment has a relatively higher $^{238}$U/$^{232}$Th content in the LS due to the addition of gadolinium elements, while the $^{238}$U/$^{232}$Th content in the LS of other experiments is all below 0.1~ppq.}
	\label{tab:LSexperiment}
\footnotesize
\renewcommand\arraystretch{1.3}
	\begin{tabular}{c|c|c|c|c}
        \hline
        & Target & Mass [t] & $^{238}$U [g/g] & $^{232}$Th [g/g] \\ \hline
        KamLAND~\cite{KamLAND:2014gul} & LS & 1,000 & (5.0$\pm$0.2)$\times$10$^{-18}$ & (1.3$\pm$0.1)$\times$10$^{-17}$ \\ \hline
        Borexino~\cite{Borexino:2017rsf} & LS & 278 & $<$9.4$\times$10$^{-20}$ & $<$5.7$\times$10$^{-19}$ \\ \hline
        SNO+~\cite{Inacio:2022vjt} & LS & 780 & (4.7$\pm$1.2)$\times$10$^{-17}$ & (5.3$\pm$1.5)$\times$10$^{-17}$  \\ \hline
        Daya Bay~\cite{DayaBay:2016ggj} & Gd-LS & 20 & (3.6$\pm$0.2)$\times$10$^{-14}$ & (2.0$\pm$0.1)$\times$10$^{-12}$ \\ \hline \hline
        JUNO expected~\cite{JUNO:2021vlw} & LS & 20,000 & 10$^{-15}$-10$^{-17}$ & 10$^{-15}$-10$^{-17}$ \\
        \hline
    \end{tabular}
\end{center}
\end{table}

The commonly employed techniques for material screening to ppq level are Neutron Activation Analysis (NAA) and Inductively Coupled Plasma Mass Spectroscopy (ICP-MS). NAA offers sensitivity comparable to ICP-MS for most materials; however, it is both more time-consuming and more expensive. Conversely, the sample preparation for ICP-MS is more complex than that for NAA. The general sensitivity of these methods is at the 10$^{-13}$~g/g level, which surpasses the radiopurity requirements in LS by several orders of magnitude. For both $^{238}$U and $^{232}$Th in LS, meticulous careful cleanliness control throughout the entire pre-treatment process is essential to enhance screening sensitivity. In this study, we employed an acid extraction method to concentrate $^{238}$U and $^{232}$Th in LS, and the concentrated samples subsequently analyzed using ICP-MS.

The rest of the paper is organized as follows: Section~\ref{sec2}  describes the experimental equipment and apparatus and provides an overview of the pre-treatment method for the LS samples. Section~\ref{sec3} presents the experimental results of the optimization of all parameters involved in the pre-treatment process of the LS samples and establishes the final pre-treatment procedure. Section~\ref{sec4}  demonstrates the $^{238}$U/$^{232}$Th recovery rates of the pre-treatment process obtained using three standard adding methods. Section~\ref{sec5} presents the results of the blank tests and the method detection limits for the entire process. Section~\ref{sec7} is the summary of this study.

\section{Experimental section}
\label{sec2}
\subsection{Instruments}
We have an ICP-MS laboratory that includes a Class 100 sample pre-treatment room, a Class 1000 cleaning room, and a Class 1000 ICP-MS measurement room. The experiment uses a ThermoFisher iCAP-Qc
Quadrupole ICP-MS instrument with a PFA concentric nebulizer, which can achieve a response of 1000-1200~counts per second (cps) for 1~ppt $^{238}$U and $^{232}$Th detection. To greatly reduce external contamination, all sample pre-treatment processes are completed within the Class 100 room, and we have developed a comprehensive cleaning procedure for the containers and instruments used in the experiments.

\subsection{Pre-treatment process}
\begin{figure}[!h]
    \centering
    \includegraphics[width=0.7\textwidth]{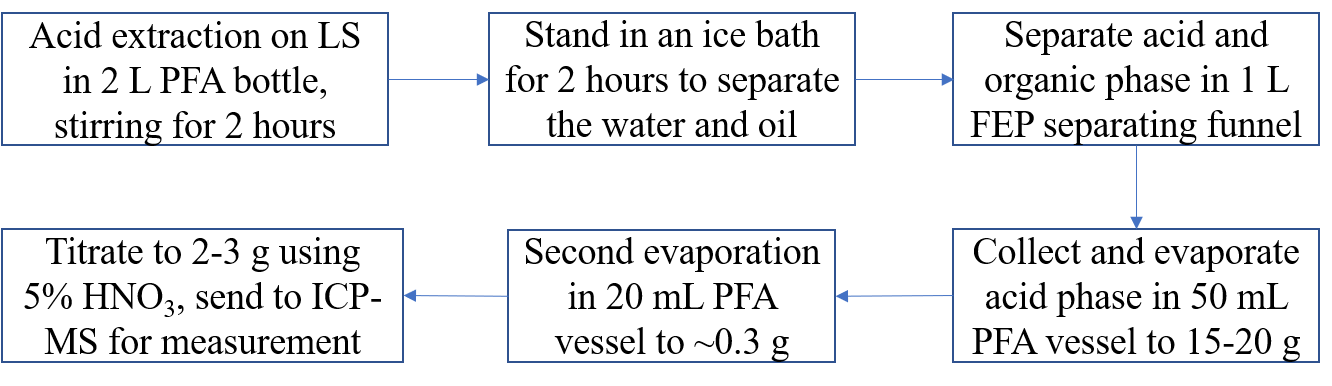}
    \caption{The entire pre-treatment process flowchart.}
    \label{fig:flowchart}
\end{figure}

Since LS is an organic phase and cannot be directly introduced into ICP-MS for analysis, it is necessary to extract and concentrate $^{238}$U and $^{232}$Th from the LS using an acidic solution before measurement by mass spectrometer. Therefore, the sample pre-treatment process primarily includes acid extraction, phase separation, and acid evaporation for enrichment. The entire pre-treatment process flowchart is shown in Fig.~\ref{fig:flowchart}.

\subsection{Labware, reagent and cleaning procedure}
All the labware and reagents used in this experiment are listed in Table~\ref{tab:reagents}. The deionized water used for cleaning and pre-treatment in the experiment is produced by the Milli-Q equipment. The IKA magnetic stirrer is used for bottle acid boiling and acid extraction stirring processes, and its built-in temperature probe can measure the actual temperature of the liquid. It is important to note that the temperature set on the instrument differs from the actual liquid temperature. For example, if the instrument is set at 135$^{\circ}$C, the measured liquid temperature using the temperature probe inserted into the liquid may be 60$^{\circ}$C. During the process of optimizing temperature parameters, we will insert the temperature probe deep into the non-sample liquid to measure the actual liquid temperature, in order to avoid contaminating the sample. The customized graphite digestion instrument is made in a honeycomb structure, and the matching PFA digestion vessel can be directly placed into the aperture for heating, covering 3/4 of the bottle depth, thus improving the efficiency of the acid chase enrichment. The $^{233}$U/$^{229}$Th standards produced by National Physical Laboratory (NPL) are stored in 2~mol$\cdot$dm$^{-3}$ nitric acid, and are used to quantify the loss of U/Th during the pre-treatment process.

\begin{table}[!h]
\begin{center}
\caption{The reagents and labware used in this work.}
\begin{threeparttable}
	\begin{tabular}{cc}
        \hline
         Reagents or Labware & Description  \\ \hline
         Ultra-pure water & Milli-Q  \\
         Heating magnetic stirrer & IKA, RH digital white \\
         Graphite digestion instrument & Customized in China\\
         Detergent & Alconox \\
         Nitric acid & OPTIMA (Fisher Scientific) \\
         Nitric acid & GENIUS \\
         Alcohol & GENIUS \\
         $^{229}$Th standard & NPL, 101.6$\pm$1.4 Bq$\cdot$g$^{-1}$ \\
         $^{233}$U standard & NPL, 1031.3$\pm$6.1 Bq$\cdot$g$^{-1}$ \\
         $^{238}$U/$^{232}$Th standard\tnote{a} & 100 $\upmu$g$\cdot$mL$^{-1}$ \\
         PFA vessel & FUJIFILM, 20~mL, 50~mL\\
         PFA bottle & OPTIMA (Fisher Scientific), 2~L  \\
         FEP separating funnel & Thermo Nalgene, 1~L \\
         Teflon coated magneton & IKA, 5~cm size \\
         \hline
	\end{tabular}
          \begin{tablenotes}
        \footnotesize
        \item[a] From national standard reference material of P.R. China.
      \end{tablenotes}
    \end{threeparttable}
	\label{tab:reagents}
\end{center}
\end{table}

All the containers are well cleaned to minimize the external contamination. Detailed treatments for 20~mL containers are shown below:

\begin{itemize}
    \item Containers with organic residues are first cleaned with 99.8\% GENIUS alcohol. All bottles are ultrasonically cleaned with 0.1\% Alconox solution and then rinsed with water.
    \item All bottles are primarily washed using TraceCLEAN equipment~\cite{traceclean} with 20\% GENIUS nitric acid for three hours at a set temperature of 220$^{\circ}$C.
    \item After rinsing with water, a secondary acid wash is performed: containers are soaked in an acid bath with a 20\% concentration of GENIUS nitric acid and left to soak overnight.
    \item A third acid wash follows, with containers being rinsed with water and then boiled in 20\% GENIUS nitric acid for 30 minutes.
    \item The radiopurity of Fisher nitric acid (5.7$\pm$2.3~ppq) is about one order better than GENIUS (70$\pm$10~ppq) based on the screening by ICP-MS. After another water rinse, a fourth acid wash is conducted, boiling the containers in 20\% Fisher nitric acid for 30~minutes. The containers are then rinsed with water, blown dry with high-pure nitrogen, and set aside for ICP-MS test.
    \item 3~mL 5\% Fisher nitric acid are added to the 20~mL containers, which are then heated and boiled for 30 minutes. After being loaded onto the instrument for measurement, the containers are considered ready for use if the ICP-MS count on $^{238}$U/$^{232}$Th is less than 10~cps.
\end{itemize}

For 50~mL PFA bottles, we exclusively use them for heating and concentrating the acid solution by approximately one order, resulting in minimal external contamination. Consequently, we limit our procedure to just the third and fourth stages of acid washing, followed by the final ICP-MS testing as previously described. In the case of 2~L PFA bottles, which are too large for the TraceCLEAN equipment and acid baths, we begin by adding roughly 90~mL of 5\% nitric acid to the bottle and agitate it. After a subsequent water rinse, we introduce 30~mL of 5\% nitric acid into the container and boil it with stirring for 30~minutes. The bottles are deemed ready for use if ICP-MS testing confirms that the $^{238}$U/$^{232}$Th concentration in the acid solution complies with the established criteria.

\section{Optimization on parameters}
\label{sec3}
One of the most critical steps in the pre-treatment process is acid extraction, which involves optimizing numerous parameters such as acid concentration, extraction ratios and repetitions, as well as stirring speed, duration, and temperature.

The effectiveness of the optimization can be reflected by the recovery efficiency. The recovery efficiency during sample pre-treatment is typically assessed by adding a known amount of a standard sample, and U/Th standards are usually preserved in acid solutions, making it difficult to uniformly mix the standard into the LS. In this study, we endeavored to introduce 0.3~g of naturally non-existent $^{233}$U and $^{229}$Th standards at a concentration of 1~ppt, stored in 5\% nitric acid and in ion form, into 10~g of linear alkylbenzene (LAB), with the expectation of achieving a concentration of approximately 30~ppq for $^{233}$U and $^{229}$Th in the LAB. Comprehensive information regarding the $^{233}$U and $^{229}$Th standards can be found in Table II.
By continuously stirring and heating until the aqueous phase completely evaporated, we obtained a relatively uniform distribution of $^{233}$U and $^{229}$Th in the LAB. Subsequently, we performed a three-stage extraction with 5\% nitric acid on this spiked LAB and completed the other steps of the pre-treatment process. The final analysis showed recovery efficiency of 104\%$\pm$10\% for $^{233}$U and 101\%$\pm$10\% for $^{229}$Th. The recovery efficiencies were close to 100\%, which we speculate is because $^{233}$U and $^{229}$Th are ion standards preserved in nitric acid solution. Their natural ionic form tends to stay in the acid phase, so even after they are distributed relatively uniformly into the LAB oil phase through stirring and heating, they easily migrate from the oil phase to the acid phase during the extraction process. However, in reality, impurities in LAB/LS may not all exist in ionic forms like the standards; therefore, we will next attempt to add 2,5-diphenyloxazole (PPO) to the LAB to further evaluate the extraction efficiency. We added PPO concentration to LAB at 103.2 g/L, and the radiopurity of PPO is screened with the similar method described in Ref.~\cite{Liu:2022srj}. $^{238}$U in PPO was detected at 0.09$\pm$0.02~ppt, and $^{232}$Th at 0.20$\pm$0.08~ppt, with the error including both the statistical error from the mass spectrometer analysis and the standard deviation among control samples. Therefore, the expected $^{238}$U and $^{232}$Th concentration in LAB solution loaded with 103.2~g/L PPO is at the level of 10~ppq. If not otherwise specified, subsequent parameter optimization experiments are based on the LAB solution loaded with 103.2~g/L PPO (PPO master solution).

\subsection{Stirring and layering}
In order to achieve a higher extraction efficiency, the acid and oil phases need to be thoroughly mixed. Experimental observation showed that using a 5~cm magnetic stir bar at a speed of 1300~rpm in a 2~L PFA bottle can thoroughly mix the water and oil phases into a white liquid. To efficiently extract $^{238}$U/$^{232}$Th from PPO master solution, we set the solution stirring time at 2~hours. After the extraction is complete, the solution is placed in an ice bath for 2~hours to accelerate the cooling and layer separation of the solution. A clear interface between the water and oil phases can be observed.

\subsection{Acid concentration and number of extraction}
To compare the extraction efficiency of different acid concentrations and number of extractions, we conducted three extractions using different concentration of nitric acid to evaluate their respective extraction performances. Extracting the PPO master solution with a high concentration (30\%) of nitric acid can lead to the precipitation of PPO~\cite{Liu:2022srj}. Therefore, we conducted extraction experiments using 5\% and 10\% nitric acid at 60$^{\circ}$C for comparison, and did not observe the precipitation of PPO in the experiments. The results are summarized in Table~\ref{tab:numExtract}. The calculation method for the extraction efficiency of each stage is to divide the amount of $^{238}$U/$^{232}$Th extracted in that stage by the total amount of $^{238}$U/$^{232}$Th extracted in the three stages. The results show that a first-stage extraction with 10\% nitric acid can achieve an extraction efficiency of over 90\% with relatively smaller Root Mean Square (RMS). Therefore, in order to save quality inspection time and improve efficiency, we have chosen a first-stage extraction with 10\% nitric acid.

\begin{table}[!h]
\begin{center}
	\caption{PPO master solution was conducted with three extractions using both 5\% and 10\% nitric acid. The calculation method for the extraction efficiency of each stage is to divide the amount of $^{238}$U/$^{232}$Th extracted in that stage by the total amount of $^{238}$U/$^{232}$Th extracted in the three stages. The average value and RMS of four parallels are also listed in the table.}
	\label{tab:numExtract}
\footnotesize
\renewcommand\arraystretch{1.3}
	\begin{tabular}{c|c|c|c|c|c|c|c}
        \hline
        \multicolumn{2}{c|}{Extraction}  & \multicolumn{2}{c|}{First-stage} & \multicolumn{2}{c|}{Second-stage} & \multicolumn{2}{c}{Third-stage} \\
        \cline{3-8}
       \multicolumn{2}{c|}{}& $^{238}$U & $^{232}$Th & $^{238}$U & $^{232}$Th & $^{238}$U & $^{232}$Th \\ \hline
        5\% nitric acid & average & 76\% & 79\% & 15\% & 16\% & 9\% & 5\% \\
        & RMS & 24\% & 26\% & 14\% & 19\% & 11\% & 8\% \\ \hline
        10\% nitric acid & average & 91\% & 93\% & 9\% & 7\% & 1\% & 1\% \\
        & RMS & 7\% & 10\% & 4\% & 7\% & 3\% & 4\% \\
        \hline
    \end{tabular}
\end{center}
\end{table}

\subsection{Acid-to-oil ratio}
In order to minimize the use of acid thus reduce external contamination in the reagent and acid-chasing enrichment process, we compared the effects of different acid-to-oil ratios (1:3 and 1:5) on the extraction of PPO master solution, and the results are summarized in Table~\ref{tab:ratioExtract}. It can be seen that the acid extraction efficiency of the two ratios (No.1 and No.2) did not show significant differences. Finally, we choose 1:5 as the acid-to-oil ratio for the acid extraction.

\begin{table}[!h]
\begin{center}
\caption{The extraction on PPO master solution was performed using different acid-to-oil ratios (1:3 and 1:5) and temperature (20$^{\circ}$C, 40$^{\circ}$C, 60$^{\circ}$C, 80$^{\circ}$C). }
	\label{tab:ratioExtract}
\footnotesize
\renewcommand\arraystretch{1.3}
	\begin{tabular}{c|c|c|c|c}
        \hline
        \multirow{2}{*}{No.} & \multirow{2}{*}{Ratio} & Temperature & $^{238}$U & $^{232}$Th \\
        & & [$^{\circ}$C] & [ppq] & [ppq] \\ \hline
        1 & 1:5  & 20 & 5$\pm$2 & 7$\pm$4 \\ \hline
        2 & 1:5  & 40 & 4$\pm$2 & 6$\pm$1 \\ \hline
        3 & 1:5  & 60 & 7$\pm$1 & 12$\pm$2 \\ \hline
        4 & 1:3  & 60 & 8$\pm$2 & 12$\pm$2 \\ \hline
        5 & 1:5  & 80 & 10$\pm$3 & 20$\pm$3 \\ \hline
    \end{tabular}
	\end{center}
\end{table}

\subsection{Temperature}
We selected four temperature points to optimize the stirring temperature: 20$^{\circ}$C, 40$^{\circ}$C, 60$^{\circ}$C, and 80$^{\circ}$C, with 80$^{\circ}$C being the heating temperature used for PPO background measurements in Ref.~\cite{Liu:2022srj}. The experimental results of using 10\% nitric acid for the extraction of the PPO master solution at an acid-to-oil ratio of 1:5 are summarized in Table~\ref{tab:ratioExtract}. It can be seen that the extraction efficiencies at 20$^{\circ}$C and 40$^{\circ}$C are significantly low. The extraction efficiency on $^{238}$U at 60$^{\circ}$C is slightly lower than that at 80$^{\circ}$C, however, the difference is larger for $^{232}$Th. Therefore, we selected 80$^{\circ}$C as the stirring temperature.

\subsection{Final procedure of pre-treatment}
According to the experience of PPO screening with the method in Ref.~\cite{Liu:2022srj}, for PPO at the 0.1~ppt U/Th level, a 1:1 acid extraction can be used 10 times with the extraction efficiency still being linear. So for the extraction of LS samples below ppq, in principle, the acid can be reused more than 200 times with an acid-to-oil ratio of 1:5. Therefore, in order to reduce reagent contamination, 150~g of acid was reused three times.

In summary, the pre-treatment parameters for 2~kg of LS are as follows:

\begin{itemize}
    \item 150~g of 10\% nitric acid is used for three extractions of the 2~kg liquid scintillator. Approximately 700~g of liquid scintillator is added to a 2~L PFA bottle for each extraction.
    \item The mixture is stirred with a 5~cm magnetic stir bar at a speed of 1300~rpm for two hours at a heating temperature of 60$^{\circ}$C.
    \item After the extraction is complete, the contents are settled for two hours in an ice bath to allow for phase separation. The majority of the top oil phase is poured out directly, while the remaining approximately 100~g of oil phase and 150~g of acid phase are separated out using a 1~L PFA separating funnel.
    \item After the water-oil phase separation, the acidic phase at the bottom is transferred to two 50~mL PFA bottles, and the acid phase is added multiple times until the liquid is heated and concentrated to 15-20~g.
    \item It is then transferred to a 20~mL PFA bottle and further heated and concentrated to a droplet (approximately 0.3~g). Finally, the volume is adjusted to 2-3~g with 5\% nitric acid for on-machine liquid analysis using ICP-MS.
\end{itemize}

\section{Recovery efficiency with optimized parameters}
\label{sec4}
Based on the optimized parameters, the final recovery efficiency was evaluated using three standard addition methods: the addition of $^{229}$Th and $^{233}$U inorganic standards, the addition of PPO, and the $^{212}$Pb loading method.

Uranium and thorium are commonly found in nature in the form of oxides, such as uranium dioxide (UO$_2$) and uranium trioxide (U$_3$O$_8$) for uranium, and thorium dioxide (ThO$_2$) for thorium. In LS, most of the U/Th impurities are typically in ion form due to the acid process used in producing LAB (nitric acid and HF acid). However, there is a risk of external contamination with U/Th oxides, which can react with the acid during pre-treatment. The specific format of U/Th impurities in organic compounds is unknown, but it is assumed to be similar to PPO. To ensure minimal loss of U/Th during the process, ion standards ($^{229}$Th and $^{233}$U) and organic standards (PPO) were employed. While $^{212}$Pb has different properties from U/Th, it is used as a standard to ensure uniform distribution in LS through radon loading in the absence of a water phase. This standard complements the use of the other two standards ($^{229}$Th and $^{233}$U in acid, PPO powder).

\subsection{$^{229}$Th and $^{233}$U standards}
Before the acid extraction, 0.3~g of 1~ppt $^{229}$Th and $^{233}$U standard samples were added to approximately 150~g of 10\% nitric acid, and 2~L of LS was extracted, undergoing all other pretreatment steps. The mass of $^{229}$Th and $^{233}$U added is equivalent to the mass of 2~kg of LS containing 10$^{-16}$~g/g of $^{238}$U/$^{232}$Th. This method was primarily used to assess the potential loss of U/Th during the acid extraction, liquid-liquid separation, and heating enrichment processes. The final results showed a $^{233}$U recovery rate of 108$\pm$11\% and a $^{229}$Th recovery rate of 115$\pm$7\%, demonstrating that the U/Th had almost no loss during the whole process. Additionally, we compared the same experiment without the addition of standard samples, and $^{229}$Th and $^{233}$U were not detected in the final liquid, indicating that the entire process was carried out in sufficiently clean containers.

\subsection{PPO addition}
We added PPO with a known $^{238}$U/$^{232}$Th content at a concentration of 2.5~g/L to 2~kg of LAB. The PPO used here and the PPO used for parameter adjustment are from the same batch, containing 0.09$\pm$0.02~ppt of $^{238}$U and 0.20$\pm$0.08~ppt of $^{232}$Th, as introduced in Section~\ref{sec3}. We did not detect any signal using the same pre-treatment method for this LAB, so we consider that the LAB itself contains a negligible amount of $^{238}$U/$^{232}$Th, and the main contamination comes from the added PPO. Based on the concentration of $^{238}$U/$^{232}$Th in the PPO and the dilution factor of the LAB, we can calculate the expected $^{238}$U/$^{232}$Th content in the liquid scintillator. By comparing the detected $^{238}$U/$^{232}$Th in the liquid after pre-treatment, we obtained a $^{238}$U recovery rate of 106$\pm$31\% and a $^{232}$Th recovery rate of 123$\pm$43\%. The large error in the results mainly comes from the uncertainty in the $^{238}$U/$^{232}$Th content of the PPO. The results show that there is no significant $^{238}$U/$^{232}$Th loss during the pre-treatment process of the liquid scintillator with $^{238}$U($^{232}$Th) content of 3(6)$\times$10$^{-16}$~g/g level.

\subsection{$^{212}$Pb loading}
The detailed method for adding $^{212}$Pb to the LS to assess the efficiency of acid extraction is described in Ref~\cite{Hu:2016jwc}. The main process involves passing nitrogen gas through a water bath and a 1200~Bq $^{220}$Rn source to carry away some of the $^{220}$Rn, and the water bath is used to increase the humidity of the nitrogen gas. Finally, the nitrogen gas is introduced into the LS. All operations are conducted in a clean glove box. The experiment showed that when nitrogen gas containing $^{220}$Rn was introduced into the LS for 74.2~hours, the $^{212}$Pb content in the LS reached equilibrium. The $\beta$-$\alpha$ cascade decay of $^{212}$Pb-$^{212}$Bi-$^{212}$Po (Bi-Po) in the LS was observed by PMTs, and the background of the entire experiment was approximately 2~counts per day.

In this experiment, we introduced nitrogen gas containing $^{220}$Rn into the LS for 3-4 days. After loading the LS, 12562~Bi-Po events were observed in a 20~mL sample measured for 50~minutes, where the half-life of $^{212}$Pb is 10.6~hours. This calculation yielded a loaded $^{212}$Pb content of 5$\times$10$^{-18}$~g/g in the LS. After acid extraction of 660~g of LS, a 20~mL sample was taken, and 49~Bi-Po events were observed using PMTs, indicating that there was still 2$\times$10$^{-20}$~g/g of $^{212}$Pb remaining in the LS. Since 24~hours had passed since the Bi-Po measurement before the acid extraction, 77\% of the $^{212}$Pb had decayed, resulting in a recovery rate of 98.2$\pm$0.3(stat.)$\pm$2(sys.)\%. This experiment demonstrates that the acid extraction efficiency of $^{212}$Pb in the LS with a content of 5$\times$10$^{-18}$~g/g is close to 100\%.

Based on the above, we have attempted three different types of standard sample loading methods: $^{229}$Th/$^{233}$U inorganic standards, PPO organic standards, and radon decay chain standards, all of which showed recovery rates close to 100\% at the level of 10$^{-16}$~g/g, with about 10\% uncertainty.

\section{Blanks and method detection limit}
\label{sec5}
\subsection{Results}
The external contamination during the sample pretreatment process is typically determined by conducting a blank experiment, which involves the entire processing and analysis procedure without adding the sample. The external contamination in this experiment mainly includes containers, reagents, and the environment. The cleaning process for the containers has been detailed in Section~\ref{sec2}, and mass spectrometer detected $^{238}$U/$^{232}$Th contamination of less than 10~cps, corresponding to 0.02~pg.

The reagents used include nitric acid and deionized water. The measurement method for 70\% Fisher nitric acid is as follows: approximately 10~g of nitric acid underwent acidification by heating for about 1.5~hours. The results of measurements for 18 parallel samples showed $^{238}$U to be 5.7$\pm$2.3~ppq and $^{232}$Th to be 5.1$\pm$1.4~ppq, with the standard deviation of the 18 parallel samples included as a systematic error. The deionized water is obtained from Milli-Q, and approximately 150~g of deionized water was heated and enriched for 5~hours. The results of three parallel samples showed $^{238}$U to be 0.38$\pm$0.04~ppq and $^{232}$Th to be 0.81$\pm$0.05~ppq, with the error also including the standard deviation of the three parallel samples. Both results also include environmental contamination during the enrichment process. In the LS experiment, 10\% nitric acid was used, and based on the measurements of Fisher nitric acid and deionized water, it was calculated that the 10\% nitric acid contained 0.17~pg of $^{238}$U and 0.22~pg of $^{232}$Th.

\begin{figure}[!h]
    \centering
    \includegraphics[width=0.5\textwidth]{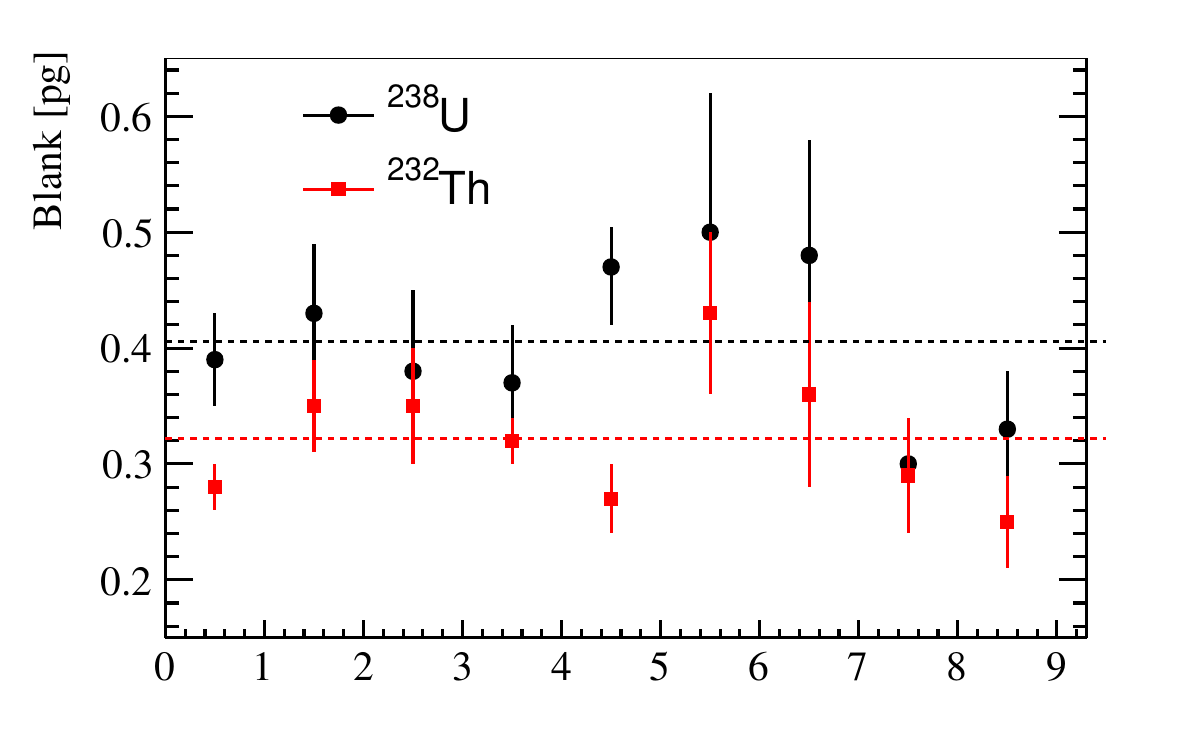}
    \caption{The amount of $^{238}$U and $^{232}$Th contamination from the blank test are shown in circle and square points. The average contamination of $^{238}$U and $^{232}$Th are shown in the black and red dash line.}
    \label{fig:blank}
\end{figure}

The results of nine blank replicates are shown in Fig.~\ref{fig:blank}. The overall experimental systematic error mainly consists of two parts. One is the mass spectrometer (accuracy 0.3~amu) and the balance (accuracy 0.1~mg), both of which make a negligible contribution and can be ignored. The other is environmental contamination in the pretreatment process, which has a certain fluctuation. The average values ($\overline{X}$) are 0.41~pg of $^{238}$U and 0.32~pg of $^{232}$Th, with standard deviation (SD) of 0.07~pg for $^{238}$U and 0.06~pg for $^{232}$Th. The method detection limit (MDL) at 99\% confidence level (C.L.) is calculated as ($\overline{X}$+$t\times$SD)/$M_{LS}$, where $M_{LS}$ is the sample mass, $t$ = 2.896 for nine replicates~\cite{MDL}. For the detection of 2~kg of LS, the MDL at 99\% C.L. is 0.30~ppq for $^{238}$U and 0.24~ppq for $^{232}$Th separately.

\subsection{Discussions}
Further improvement of the detection limit can be achieved by enriching more LS, provided that external contamination does not increase linearly with the amount of sample enrichment. External sources of contamination include containers, reagents, and environmental pollutants throughout the entire process. The containers used for pretreatment remain unchanged, with only an increase in the number of enrichments, thus the container contamination remains constant. The acid extraction reagent uses only 10\% nitric acid. According to the PPO extraction experience, the acid and PPO (0.1~ppt U/Th) can still detect the linear extraction of U/Th after being reused 10 times at a 1:1 ratio. Therefore, for the extraction of LS containing U/Th at the level of 0.3~ppq, in principle, the acid used for LS extraction in a 1:5 ratio can be reused nearly 60 times. Therefore, 150~g of 10\% nitric acid can be reused 30 times for the extraction of 20~kg of LS, with a tenfold increase in sample volume and no change in reagent background. Only the final acid chase enrichment step of the entire operation is exposed to long-term (several hours) air contamination. However, as the amount of acid chase remains constant, this contamination also remains unchanged. The liquid-liquid separation step involves short-term exposure to air (on the order of minutes), which is minimal compared to the time for the acid chase.

We attempted a blank experiment of acid extraction 15 times for nearly one week, equivalent to extracting 11~kg of LS with 150~g of nitric acid. The average U/Th results of the three parallel samples were 0.3~pg, with a standard deviation of 0.1~g, while the average U/Th of blanks in Fig.~\ref{fig:blank} is 0.3-0.4~pg for three times extraction. This indicates that the blank extraction with acid did not linearly increase with the number of extractions, demonstrating that the detection limit can be improved by enriching more LS. In summary, increasing the enrichment to 20~kg will not significantly increase the blank linearly, and is expected to improve the detection limit to 0.03~ppq, while the acid extraction time in the pretreatment process will increase from 2 days to 15 days.

\section{Summary}
\label{sec7}
This study developed a pre-treatment method for LS samples based on ICP-MS. The method uses acid extraction to enrich and concentrate the $^{238}$U/$^{232}$Th content in the LS, resulting in a detection limit of 0.30 (0.24)~ppq for $^{238}$U ($^{232}$Th) in 2~kg of LS. In terms of recovery efficiency, we tested three standard addition methods to evaluate the loss of U/Th in the LS during the pre-treatment process. These methods included the addition of naturally non-existent $^{229}$Th and $^{233}$U, quantitative addition of known concentrations of PPO, and loading of radon progeny $^{212}$Pb, all of which confirmed that the recovery efficiency of $^{238}$U/$^{232}$Th in the LS during the pre-treatment process approaches 100\%, subject to a 10\% uncertainty. Enriching more samples is expected to further improve the detection limit.

\section*{Acknowledgements}
This work is supported by the Youth Innovation Promotion Association of the Chinese Academy of Sciences, CAS Project for Young Scientists in Basic Research (YSBR-099), and National Natural Scienee Foundation of China under Grant No.12125506.

\bibliographystyle{unsrt}
\bibliography{reference}

\end{document}